\documentclass{icrc}
\usepackage{times}
\usepackage{graphicx} 
\def\lesssim{\mathrel{\hbox{\rlap{\hbox{\lower4pt\hbox{$\sim$}}}\hbox{$<$}}}}
\def\gtrsim{\mathrel{\hbox{\rlap{\hbox{\lower4pt\hbox{$\sim$}}}\hbox{$>$}}}}

\newcommand{\ffrac}[2]
  {\left( \frac{#1}{#2} \right)}
\begin{document}
\title{Arrays is Space to detect Upward ${\tau}$ and Highest Altitude Showers}


\author{D.Fargion}
\affil{ Physics Department and INFN Rome University 1,
     Pl.A.Moro 2, 00185, Rome, Italy}

\correspondence{daniele.fargion@roma1.infn.it}

\firstpage{1} \pubyear{2001}

\maketitle
\begin{abstract}
Ultra High Energy, UHE, upward Tau neutrinos $\nu_{\tau}$,
$\bar\nu_{\tau}$, above hundred TeVs and up to tens PeV energies,
of relevant astrophysical nature, may lead to $UHE$  Taus  and
consequent Up-ward Tau air-Showers (\textbf{UPTAUS}) after
interaction on Earth crust surface. The UPTAUS discover may open
a new UHE Tau Neutrino Astrophysics. A new generation of Gamma,
X, optical and Radio Arrays in Space may discover, in the same
Auger spirit, such up-coming Air-showers as well as an additional
Tau signal: the nearly Horizontal Tau AIR-Shower
(\textbf{HORTAUS}) originated by $UHE$ neutrinos $\tau$ at
$10^{19} eV$ energies arising from a thin Earth crust corona at
few tens of degree below the horizons; a degree above the
horizons, there should be over common diffused cosmic ray albedo,
an additional High Altitudes (nearly Horizontal) Showers
(\textbf{HIAS}), by more common Cosmic Rays primaries at PeVs up
to EeV and ZeV energies, both of hadronic or of electro-magnetic
$\gamma$ nature. Mini-arrays detectors in high Altitude Balloons
tails facing the horizons and the Earth below may also reveal
both UPTAUS, and HIAS, and more rarely HORTAUS. Gamma Burst,
Cherenkov flashes and rarer muons observed by  high mountains
peak arrays or high quota balloon array may better probe UPTAUS
and HORTAUS. Present and future X-Gamma satellites as Beppo-Sax
may be also able to discover HORTAUS and HIAS discovering
transient events by UHE source (as the Crab) while in occultation
by Earth. Because of the large shower distances and the Earth
magnetic field , HIAS and HORTAUS showers will be split in a
Twin-Beams fan-shaped  wide arc, orthogonal to $B_\oplus$, with a
strong azimuths modulation. Gamma bursts and rare muons or
neutrons at the horizon, associated with HIAS, HORTAUS may be
discovered by Glast and AMS satellite.
\end{abstract}

\section{Introduction: Downward  Air-Showers}
Downward  Air-Showers have been studied in last century leading
to new advances in cosmic rays spectra and compositions as well
as to highest energy particle physics. Cosmic Rays studies
clarified fundamental physics as the matter and anti-matter
(electron-positron) or the pion-muon interaction nature opening
the gate toward present quark and lepton understanding of
elementary particles. Air Showers are commonly downward chain
reactions originated by incoming  cosmic rays (nucleons, nuclei,
gamma) on upper Earth Atmosphere. Earth itself absorbs most
upcoming cosmic rays, suppressing also PeVs upward muons
secondaries produced by noisy Atmospheric $\nu_{\mu}$, $
\bar\nu_{\mu}$ Neutrinos. Downward Air-Showers, hadronic or
electro-magnetic, have been widely modeled and measured.  At first
sight Up-ward Air-showers are totally forbidden by the severe
opacity of the Earth below both to muons as well as (by
terrestrial growing shadowing) also to $UHE$ neutrinos,
$\nu_{e}$, $\bar \nu_{e}$, $\nu_{\mu}$, $\bar \nu_{\mu}$ at tens
TeVs energies and above. The surviving upward $GeVs-TeVs$ muons,
founded in AMANDA, Baikal, MACRO detectors are \textit{not}
usually, able to lead to any upward Air showers. Moreover UHE
$\nu_{e }$, $ \bar\nu_{e }$ will produce very confined shower
(the LPM, Landau, Pomeranchuk, Migdal effect) within a very thin
earth layer (less than few meters) immediately buried in there;
upward UHE $ 10^{13}\div 10^{14} eV $, atmospheric or
astrophysical $\nu_{\mu}$, $\bar \nu_{\mu}$ and their muons
produced on a thin earth crust (few Kms depth)  will freely
escape from the Earth with very rare catastrophic bremsstrahlung
showering in the atmosphere. UHE $\nu_{\tau}$, $\bar \nu_{\tau}$
have been usually neglected due to the Tau short life-time.
Moreover one should remind that UHE $\nu_{\tau}$,
$\bar\nu_{\tau}$ are rarely ($<10 ^{-5}$) produced in high
energy  astrophysical sources. However the recent discover by
Super Kamiokande of a full flavor mixing $\nu_{\mu}$ $\nu_{\tau}$
implies a nearly homogeneous abundance of all UHE $\nu_{e }$,
$\nu_{\mu}$,$\nu_{\tau}$ flavors (Fargion 1997, 1999).
\section{UPTAUS: The Upward  Tau Air-shower}
 The Upward  UHE $\nu_{\tau}$, being less opaque to the Earth, it may
interact within the terrestrial surface crust (nearly Kms depth)
leading to $ 10^{14}\div 10^{16}eVs $ events in Tau Air-Shower
UPTAUS form (Fargion 2000). Correlation between UPTAUS and
observed Terrestrial Gamma Flashes (TGF) has been noted (Fargion
2000,2001, HE2.5). Here we present and discriminate the three
main new  kind of Air-Showers to be detectable on balloons or
satellites.. (1) Upward $\tau$ Air-Shower UPTAUS around PeVs
energies. (2) The nearly horizontal HORTAUS within $ 10^{17}-
10^{19} eV$ energies upcoming from the horizons edges.(3) A high
altitude Air-Showers originated by UHECR at the edges of GZK
energies, HIAS. The nature of UPTAUS and HORTAUS are analogous,
but at different energy windows: $10^{15}$- $10^{15}$eV ,
$10^{19}$ eV respectively. Their bang in matter and bang in air
remind the \textit{Tau Double Bang} (Learned,Pakvasa 1995). They
are different respect downward air-showers. Indeed the $\tau$
decay is bounded to a limited size and has an hybrid
electro-magnetic and hadronic interaction with matter. The $\tau$
main air-shower channels must reflect the rich and structured
behaviours of $\tau$ decay modes. Let us label the main UHE decay
channels (hadronic or electro-magnetic) and the consequent
air-shower
imprint, with corresponding probability ratio in following Table. 
\begin{figure}[h]
\includegraphics[width=0.45\textwidth]{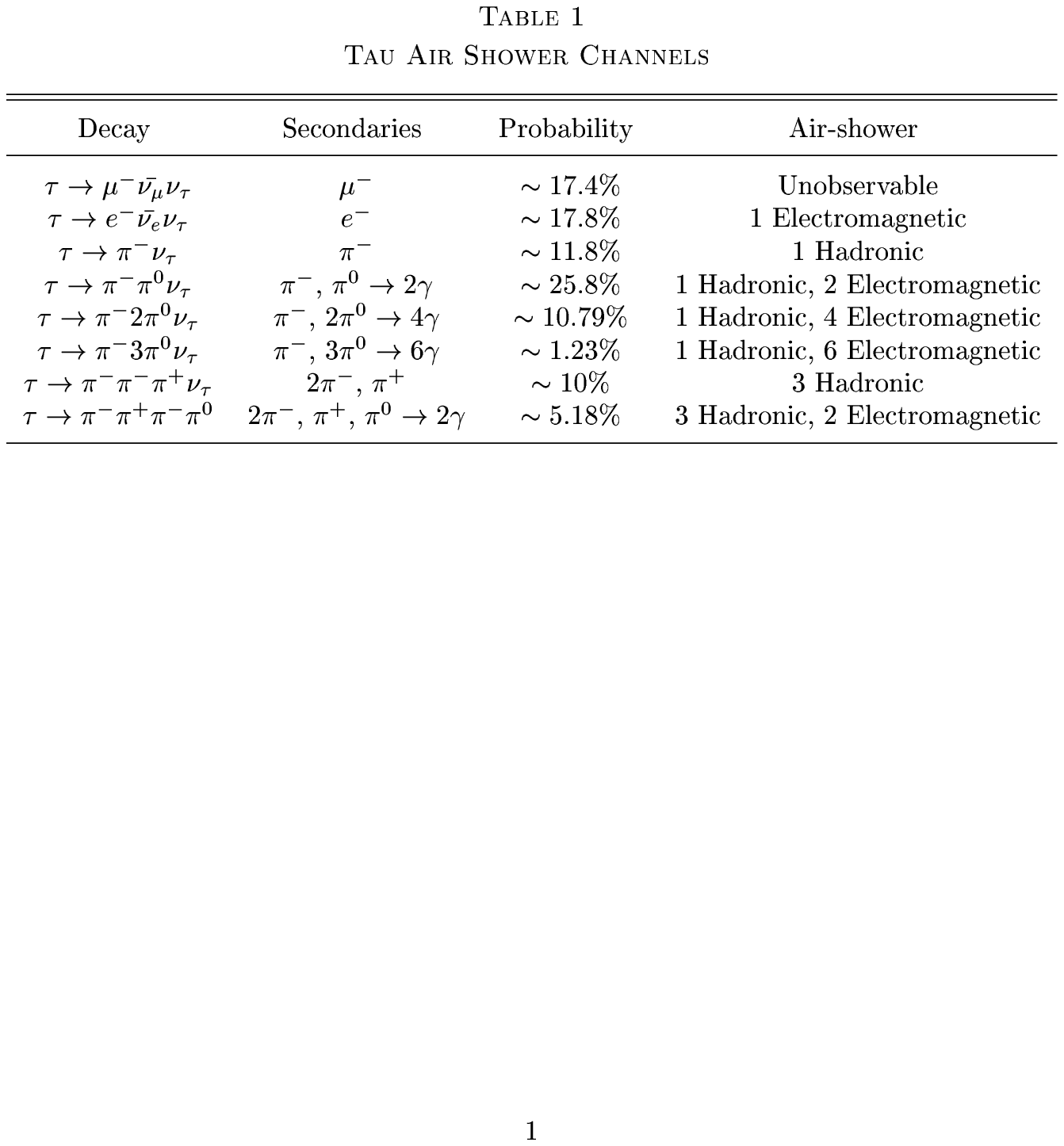} %
\caption{The Table above shows  the characteristic Tau Air-Shower
originated by an Ultra High Energy neutrino. The decay channels
are ruled by elementary particle Physics, leading to different
Shower composition: Hadronic, Electro-Magnetic or an hybrid one.
These signatures  should be verified by a large set of UPTAUS and
HORTAUS. The high atmosphere density on Sea level makes UPTAUS
poor source of UHE surviving muon secondaries; also HORTAUS
originated from a mountain chain are  muon-poor. On the contrary
HORTAUS ans HIAS  originated at high quota may be muon rich}
\end{figure}
\begin{figure}[bt]
\begin{center}
\includegraphics[width=0.45\textwidth]{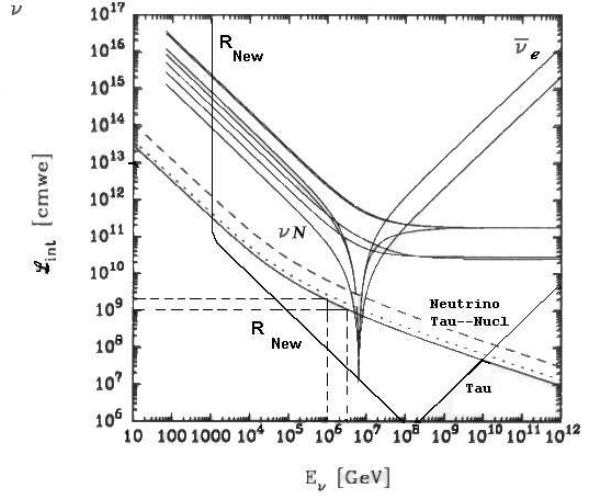}
\end{center}
  \caption {The UHE neutrino ranges  $\nu_{\tau} N$  as a function of UHE neutrino energy in Earth with
    overlapping the resonant $\bar{\nu}_e e$, interactions (Gandhi et all
    1998)and $\tau$ ranges (Fargion ICRC 1999,2001). $R_{New}$ shows the interaction length due to New physics for rock $\rho =3$}.
    \label{fig:boxed_graphic 5}
\end{figure}
The UPTAU cascade occur at high quota where  atmosphere is less
dense  and imply longer final shower tails. For this reasons the
scale time of the $\gamma$ are longer (hundreds microseconds) and
harder than downward ones. The UPTAUS maybe therefore bent in a
thin-twin fan arcs few degree size by Earth magnetic field leading
to a twin  shower beams in  deflected (positive-negative) thin
elliptical cones. The deflection has a strong azimuth and zenith
imprint as well as a characteristic geological signature (by Earth
density). It is negligible observing to North and South, while it
is maximal observing East and West. Deflection and opening angles
are maximal in Asia where terrestrial magnetism is largest as
indeed observed in TGFs data (Fargion 2000). If the Twin-Beam
UPTAUS is observed by a wide spaced array in Space its spatial and
temporal (Fargion 2000,2001) arrival structure should be
observed. The $\nu_{\tau}$, $\bar\nu_{\tau}$--$\tau$ HORTAUS, may
be generated efficiently also in front of a mountain chains :
their detection on ground or from the mountain arrays has been
discussed  (Fargion et all ICRC 26,1999; Fargion ICRC 27,HE 2001).
The expression that describe the probability to observe a UPTAUS
and HORTAUS Shower by an UHE $\nu_{\tau}$, $\bar\nu_{\tau}$  is
suppressed by Earth opacity and it is proportional to the
probability to make a $\tau$ within the Earth (or mountain) crust:
\begin{equation}
P(\theta,\, E_{\nu}) = e^{\frac{-2R_\oplus \sin
\theta}{R_{\nu_{\tau}}(E_{\nu})}} (1 - e^{-
\frac{R_{\tau}(E_{\tau})}{R_{\nu_{\tau}}(E_{\nu})}}) \, .
\end{equation}
Here $\theta$ is the angle below the horizons while standing on
the Earth (see Fargion 2001). From height $h_{2}$ (for instance
the satellite orbit),the same expression becomes:
$P(\theta,\, E_{\nu}) = $
\begin{equation}
e^{\frac{-2\sqrt{(R_\oplus+h_2)^2 \sin^2 \theta -[(R_\oplus+h_2
)^2-R_\oplus^2]}}{{R_{\nu_{\tau}}(E_{\nu})}}}(1 - e^{-
\frac{R_{\tau}(E_{\tau})}{R_{\nu_{\tau}}(E_{\nu})}}) .
\end{equation}
 where $\theta$ from height $h_2$ reaches a minimal values when
  it is just tangent to the Earth: then $\theta_{2 min}$ is:
 \begin{equation}
\theta_{2 min}=
\frac{\sqrt{(R_\oplus+h_2)^2-(R_\oplus)^2}}{(R_\oplus)}
\end{equation}
This angle from satellites at height $h_2$ $=500 Km$ is 22.4
degrees below the horizons.
\begin{figure}[h]
\begin{center}
\includegraphics[width=0.45\textwidth]{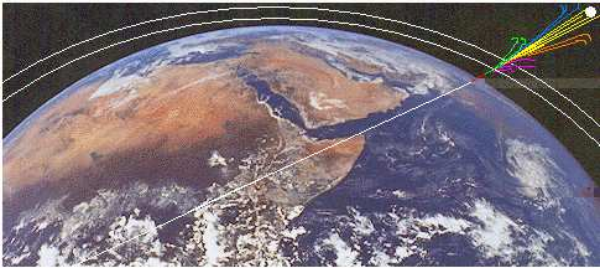} %
\caption{The Upward Tau Air-Shower $UPTAUS$ originated by UHE
$\nu_{\tau}$ (white track) crossing the Earth and interacting
within the thin Earth Crust layer. The UHE $\tau$ track (red line)
at PeVs decay in air leading to a complex shower: a twin-split
green and violet short spirals stand respectively for positron
and electron fan-shaped components; longer twin-split blue-orange
tracks describe the positive-negative muons. Final surviving
Bremsstrahlung gamma, X and Cherenkov photons are flashing in a
twin wide, but thin, yellow fan-shaped beams to satellite.}
\end{center}
\end{figure}
The visible Earth surface from a satellite, like BATSE, at height
$h \sim 400$ Km and the consequent effective volume for UHE
$\nu_{\tau} N$ PeVs interaction and $\tau$ air shower beamed
within $\Delta \Omega \sim 2 \cdot 10^{-5} rad^2$ is: (note
$<\rho> \simeq 1.6$ because 70 \% of the Earth is covered by seas)
The effective volume and the event rate should be reduced, at
large nadir angle ($\theta > 60^o$) by the atmosphere depth and
opacity (for a given $E_{\tau}$ energy). Therefore the observable
volume may be reduced approximately to within $15 Km^3$ values and
the expected UHE PeV event rate is (Fargion 2000)
\begin{equation}
  N_{ev} \sim 150 \ffrac{h}{400} \frac{\mathrm{events}}{\mathrm{year}}    \qquad (E_\tau \sim 3 \mathrm{PeV})
\end{equation}
It is well possible that UPTAUS have been already observed
(Fargion 2000,2001) as upward Terrestrial Gamma Flashes. Present
and future satellite as Glast should be able to confirm the
UPTAUS discover.
\section{HIAS :The High Altitude air-Shower}
Earth atmosphere is continuously reflecting a noisy cosmic ray
albedo due mainly to neutral pion decays in gamma upward; these
pions are secondaries of common incoming cosmic rays hitting the
Earth atmosphere. Their low energy statistical overlapping is not
source of any sudden showers. However the detection of UPTAUS and
in particular of HORTAUS  from  satellite is disturbed by a
different upcoming competitive shower signal:
  the HIAS, the High Altitude, nearly tangential, UHE cosmic
  ray,  Showers. Indeed the atmosphere depth at
  slant depth greater than $2 10^{3}$ $g cm^{-2}$
  is commonly opaque (suppression $10^{-3}$) to most cosmic rays showers (but not to muons ones).
  Then UHECR cannot send signals upward while crossing all
  the Earth dense atmosphere.  On the contrary  at very high altitude air become so diluted that it may become
again transparent to horizontal UHECR. At more high quota the
paucity of air makes rare the UHECR interactions and  there are
not showers again. There  is an optimal height which permits to
tangential UHECR to lead to powerful showers. The determination
of this height is not obvious. The author is not aware of any
derivation of this height value. Here we derive a new general
expression. The geometry of the problem needs a simple diagram.
\begin{figure}[h]
\begin{center}
\includegraphics[width=0.45\textwidth]{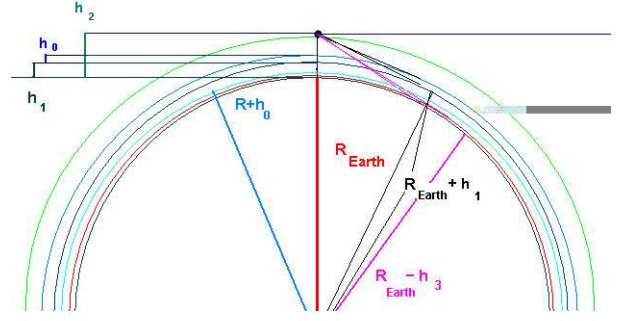}
\caption{The figure above  shows the trajectories for nearly
Horizontal Tau Air-Shower HORTAUS, by an UHR $\tau$ made by UHE
$\nu_{\tau}$ at $10^{19}$ eV energies (white track) crossing the
Earth crust at depth $h_3$ (few Km) and interacting with  tens or
a hundred terrestrial Km layers. The same upcoming HORTAUS hits
the observer (mountains,balloons,satellite) at height $h_2$. HIAS
produce at a characteristic quota $h_1$ a long track shower
within a thin corona  height $h_0$, (defined by density decay
law), reaching the observer (balloons,satellite) at quota $h_2$.}
\end{center}
\end{figure}
\begin{figure}[h]
\begin{center}
\includegraphics[width=0.45\textwidth]{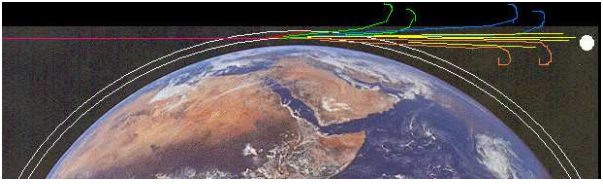} %
\caption{The nearly horizontal High Altitude Showers, $HIAS$.
Their Hadronic or electro-magnetic interaction defines the
altitudes where the shower is originated and propagate. The
terrestrial magnetic field leads to electron pairs and muon pairs
splitting and a very wide twin gamma beam in a fan shape, upward
and downward.}
\end{center}
\end{figure}
The properties of HORTAUS are comparable to UPTAUS but they are
bounded in a thin terrestrial layer and at a higher (EeV) energy
band. The height $h_0$ is the characteristic exponential air
density height, $h_1$ is the last transparent air quota where High
Altitude Air-Shower take place, $h_2$ is the satellite or balloon
quota, $h_3$ is the deepest distance from the Earth surface for
Horizontal HORTAUS. We first estimate the atmosphere depth
crossed horizontally:
\begin{equation}\label{3}
 X =  \int_{- \infty}^{\sqrt{(R_\oplus+h_2)^2-(R_\oplus+h_1)^2}}
     n_0 e^{-h/h_0} dx
\end{equation}
 We then infer that this slant depth is comparable to common  vertical one.
  This request defines  and fix the fine tuned height $h_1$
  value. Because for satellites the distances are large we may
  approximate above integral in:
\begin{equation}\label{1}
  X = \int_{- \infty}^{+ \infty} n_0 e^{-h/h_0} dx \equiv n_0 h_0
\end{equation}

\begin{equation}\label{2}
  \frac{X}{n_0} \simeq  2 \int_{0}^{+ \infty} e^{-h_1/h_0} e^{-\frac{x^2}{2(R+h_1)h_0}}
   dx \cong h_0
\end{equation}
The consequent trascendental equation that fix the height $h_1$ is
\begin{equation}\label{4}
  2 \sqrt{\pi} e^{-h_1/h_0} \sqrt{(R_\oplus+h_1)h_0} =h_0
\end{equation}
\begin{equation}\label{5}
\frac{h_1}{h_0} =\frac{3}{2} \ln 2 + \frac{1}{2} \ln \left(
\frac{R_\oplus+h_1}{h_0} \right) - \ln A + \frac{1}{2} \ln  \pi
\end{equation}
 Where  $h_0$= $8,55$ Km, $R_\oplus$ is the Earth radius, A is an additional
 parameter of order of unity that calibrate the shower slant
 depth to its incoming energy as well as to its hadronic or
 electro-magnetic interaction nature.
  After simple substitution  and applying known empirical laws for
   the logarithmic growth of the X slant depth one derives respectively for hadronic and gamma UHECR showers:
\begin{equation}\label{5}
 A_{Had.}=0.792 \left[1+0.02523 \ln\ffrac{E}{10^{19}eV}\right]
\end{equation}
\begin{equation}\label{5}
 A_{\gamma}=\left[1+0.04343\ln\ffrac{E}{10^{19}eV}\right]
\end{equation}
The corresponding solution for the trascendental equations above
are:
\begin{equation}\label{5}
{h_{1_{Had.}}} =44.08- 8.55\ln[1+0.02523 \ln\ffrac{E}{10^{19}
eV}] \; $Km$
\end{equation}
\begin{equation}\label{5}
{h_{1_{\gamma}}} =42.09 - 8.55\ln[1+0.04343 \ln \ffrac{E}{10^{19}
eV}] \; $Km$
\end{equation}
The characteristic distances crossed from this altitudes to the
satellites are hundred of Kilometers. Therefore there is an
important role of the magnetic fields in bending the HIAS
trajectories in a fan-shaped arc. This allow to observe a HIAS
not only at its origination direction (nearly one degree above
the Earth edge, from a satellite) but also below the Earth
horizons. Obviously the down-ward component of the HIAS may reach
back the Earth but with a negligible strength. Therefore, for the
same reason, there are not HIAS appearing above (one degree) from
the Earth horizons observing on satellites. On the contrary HIAS
may hit high quota baloons ($20-30 Kms$) quota both mainly above
the terrestrial horizons and rarely nearly horizontally and more
rarely below the horizons. Therefore baloons (possibly in arrays)
are ideal detectors for HIAS.  The upcoming HIAS are also
important because they may be confused as a HORTAUS from
satellites, but not in mountain observatories. The trajectory and
the morphology of the two events maybe discriminate. HIAS while
being able to appear below the horizons from satellite, cannot
completely mimic  UPTAUS because their track, spectra, timing are
quite different.
\section{HORTAUS :The Horizontal Tau Air-shower }
The possibility to UHECR neutrino to slip tangent to Earth crust
at optimal neutrino interaction length and longest Tau range makes
them ideal signals as HORTAUS. Their maximal probability occurs
as shown in Fig2, at the crossing of Tau-Neutrino Lengths near
$E_{\nu_{\tau}}$ = $10^{19} eV$, near GZK cut off
bounds(Fargion,Mele,Salis 1999) where neutrino interaction lengths
is comparable to the Tau one. At the same energy range one may
find the characteristic height, or negative depth, $h_3$ few Km,
where the HORTAUS should begin to make a UHE Tau; consequently Tau
shower should   deploy  neither at too low atmosphere (being
absorbed), nor at too high atmosphere (where no shower may be
amplified). HORTAUS may also turn upward by geo-magnetic fields.
In particular from a mountain HORTAUS within$ 10^{16-19} eV$ may
appear at angles below the original ones leading to more frequent
(apparent) UPTAUS events; their timing and spectra and morphology
is different from direct PeVs UPTAUS. The height where Tau decay
which define also the optimal Tau energy equation for HORTAUS is
analogous, but more complex than the HIAS one.
\begin{figure}[h]
\begin{center}
\includegraphics[width=0.45\textwidth]{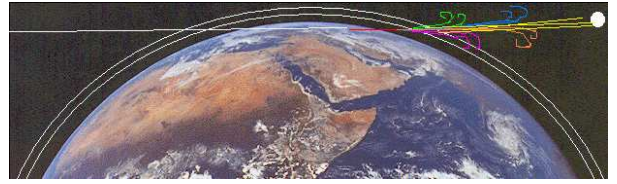}
\caption{The  Horizontal Tau Air-Shower $HORTAUS$ originated by
an Ultra High Energy neutrino at $10^{19} eV$ energies (white
track) crossing the Earth and interacting within Earth Crust
layer. The UHE Tau track (red line) decay in air leading to a
longer complex shower comparable to the described ones above}
\end{center}
\end{figure}
The trascendental equation that defines the Tau distance $c\tau$
analogous to the HIAS one are:
\begin{equation}\label{13}
  \int_{0}^{+ \infty} n_0 e^{-\frac{\sqrt{(c\tau+x)^2+R_\oplus^2} - R_\oplus}{h_0}}
   dx \cong n_0 h_0 A
\end{equation}
\begin{equation}\label{14}
  \int_{0}^{+ \infty} n_0 e^{-\frac{(c\tau+x)^2} {2h_0R_\oplus}}
   dx \cong n_0 h_0 A
\end{equation}
\begin{equation}\label{15}
 c\tau = \sqrt{2R_\oplus h_0}
 \sqrt{ln \ffrac{R_\oplus}{c\tau} - ln A }
\end{equation}
Where $A=A_{Had.}$ or $A=A_{\gamma}$ are the same defined in
eq.$10-11$. The solution of this equation leads to a
characteristic UHE $c\tau_{\tau}$ = $546 \;Km$ decay distance  at
height $h= 23$ Km where the HORTAUS begins to shower. This imply
a possibility to discover efficiently by satellite and ballons
arrays  UHE $\nu_{\tau}$, $\bar\nu_{\tau}$ up to $1.11$
$10^{19}eV $. Other consequences of the HORTAUS, UPTAUS and HIAS
physics will be shown soon elsewhere.  Arrays in valleys,
mountains, airlines, ballons and satellites should easily
discover their traces.

\begin{acknowledgements}
The author thanks Dr.Andrea Aiello  for useful discussions and
P.G. De Sanctis Lucentini for numerical tests and corrections.
\end{acknowledgements}


\end{document}